\documentstyle[epsfig,12pt,a4p]{article}

\newcommand{\pipipi}{\mbox{$\pi^+\pi^-\pi^0$ }}

\newcommand{\kstkst}{\mbox{$K^*(892) \overline K^*(892)$ }}

\newcommand{\pipipipi}{\mbox{$\pi^+\pi^-\pi^+\pi^-$ }}
\newcommand{\pipi}{\mbox{$\pi^+\pi^-$ }}
\newcommand{\kkpi}{\mbox{$K^{0}_{S} K^{\pm} \pi^{\mp}$} }
\newcommand{\pipig}{\mbox{$\pi^+\pi^-\gamma$ }}
\newcommand{\rhog}{\mbox{$\rho^0\gamma$ }}
\newcommand{\phig}{\mbox{$\phi\gamma$ }}
\newcommand{\kkg}{\mbox{$K^+K^-\gamma$ }}
\newcommand{\etapipi}{\mbox{$\eta \pi^{+} \pi^{-}$} }
\begin{document}
\begin{titlepage}
\def\footnoterule{\hrule width 1.0\columnwidth}
\begin{tabbing}
put this on the right hand corner using tabbing so it looks
 and neat and in \= \kill
\> {30 September 1998}
\end{tabbing}
\bigskip
\bigskip
\begin{center}{\Large  {\bf A measurement of the branching fractions of the
$f_1(1285)$ and $f_1(1420)$
produced in central pp interactions at~450~GeV/c}
}\end{center}
\bigskip
\bigskip
\begin{center}{        The WA102 Collaboration
}\end{center}\bigskip
\begin{center}{
D.\thinspace Barberis$^{  4}$,
W.\thinspace Beusch$^{   4}$,
F.G.\thinspace Binon$^{   6}$,
A.M.\thinspace Blick$^{   5}$,
F.E.\thinspace Close$^{  3,4}$,
K.M.\thinspace Danielsen$^{ 11}$,
A.V.\thinspace Dolgopolov$^{  5}$,
S.V.\thinspace Donskov$^{  5}$,
B.C.\thinspace Earl$^{  3}$,
D.\thinspace Evans$^{  3}$,
B.R.\thinspace French$^{  4}$,
T.\thinspace Hino$^{ 12}$,
S.\thinspace Inaba$^{   8}$,
A.V.\thinspace Inyakin$^{  5}$,
T.\thinspace Ishida$^{   8}$,
A.\thinspace Jacholkowski$^{   4}$,
T.\thinspace Jacobsen$^{  11}$,
G.T\thinspace Jones$^{  3}$,
G.V.\thinspace Khaustov$^{  5}$,
T.\thinspace Kinashi$^{  13}$,
J.B.\thinspace Kinson$^{   3}$,
A.\thinspace Kirk$^{   3}$,
W.\thinspace Klempt$^{  4}$,
V.\thinspace Kolosov$^{  5}$,
A.A.\thinspace Kondashov$^{  5}$,
A.A.\thinspace Lednev$^{  5}$,
V.\thinspace Lenti$^{  4}$,
S.\thinspace Maljukov$^{   7}$,
P.\thinspace Martinengo$^{   4}$,
I.\thinspace Minashvili$^{   7}$,
T.\thinspace Nakagawa$^{  12}$,
K.L.\thinspace Norman$^{   3}$,
J.P.\thinspace Peigneux$^{  1}$,
S.A.\thinspace Polovnikov$^{  5}$,
V.A.\thinspace Polyakov$^{  5}$,
V.\thinspace Romanovsky$^{   7}$,
H.\thinspace Rotscheidt$^{   4}$,
V.\thinspace Rumyantsev$^{   7}$,
N.\thinspace Russakovich$^{   7}$,
V.D.\thinspace Samoylenko$^{  5}$,
A.\thinspace Semenov$^{   7}$,
M.\thinspace Sen\'{e}$^{   4}$,
R.\thinspace Sen\'{e}$^{   4}$,
P.M.\thinspace Shagin$^{  5}$,
H.\thinspace Shimizu$^{ 13}$,
A.V.\thinspace Singovsky$^{ 1,5}$,
A.\thinspace Sobol$^{   5}$,
A.\thinspace Solovjev$^{   7}$,
M.\thinspace Stassinaki$^{   2}$,
J.P.\thinspace Stroot$^{  6}$,
V.P.\thinspace Sugonyaev$^{  5}$,
K.\thinspace Takamatsu$^{ 9}$,
G.\thinspace Tchlatchidze$^{   7}$,
T.\thinspace Tsuru$^{   8}$,
M.\thinspace Venables$^{  3}$,
O.\thinspace Villalobos Baillie$^{   3}$,
M.F.\thinspace Votruba$^{   3}$,
Y.\thinspace Yasu$^{   8}$.
}\end{center}

\begin{center}{\bf {{\bf Abstract}}}\end{center}

{
\par
A study of the $f_1(1285)$ and $f_1(1420)$
produced in central pp interactions has been performed.
For the first time in a single experiment the branching fractions
of both mesons in all major decay modes have been determined.
Both the $f_1(1285)$ and $f_1(1420)$ are consistent with being produced
by double Pomeron exchange.
}
\bigskip
\bigskip
\bigskip
\bigskip\begin{center}{{Submitted to Physics Letters}}
\end{center}
\bigskip
\bigskip
\begin{tabbing}
aba \=   \kill
$^1$ \> \small
LAPP-IN2P3, Annecy, France. \\
$^2$ \> \small
Athens University, Physics Department, Athens, Greece. \\
$^3$ \> \small
School of Physics and Astronomy, University of Birmingham, Birmingham, U.K. \\
$^4$ \> \small
CERN - European Organization for Nuclear Research, Geneva, Switzerland. \\
$^5$ \> \small
IHEP, Protvino, Russia. \\
$^6$ \> \small
IISN, Belgium. \\
$^7$ \> \small
JINR, Dubna, Russia. \\
$^8$ \> \small
High Energy Accelerator Research Organization (KEK), Tsukuba, Ibaraki 305,
Japan. \\
$^{9}$ \> \small
Faculty of Engineering, Miyazaki University, Miyazaki, Japan. \\
$^{10}$ \> \small
RCNP, Osaka University, Osaka, Japan. \\
$^{11}$ \> \small
Oslo University, Oslo, Norway. \\
$^{12}$ \> \small
Faculty of Science, Tohoku University, Aoba-ku, Sendai 980, Japan. \\
$^{13}$ \> \small
Faculty of Science, Yamagata University, Yamagata 990, Japan. \\
\end{tabbing}
\end{titlepage}
\setcounter{page}{2}
\bigskip
\par
\par
In central production the $J^{PC}$~=~$1^{++}$
$f_1(1285)$ and $f_1(1420)$ are clearly
observed~\cite{wa76kkpi,cenkkpi}.
In contrast,
there is no evidence for any $0^{-+}$ contribution in the 1.28 and 1.4~GeV
regions~\cite{wa76kkpi,cenkkpi,cenetapipi}.
This suppression of $0^{-+}$ states in central production
was first observed
by experiment WA76 at the CERN Omega Spectrometer~\cite{wa76kkpi}
and has been independently confirmed by the
E690 experiment at Fermilab~\cite{E690}.
Although the reason for this
suppression of $0^{-+}$ states is not known it has a very important
implication since it can help us determine more
precisely the characteristics of the
$f_1(1285)$ and $f_1(1420)$ than is possible in other experiments
which see both $0^{-+}$ and $1^{++}$ states.
\par
Experiment WA102 is designed to study centrally produced
exclusive final states formed in
the reaction
\begin{equation}
pp \rightarrow p_{f} (X^0) p_{s}
\label{eq:a}
\end{equation}
at 450 GeV/c.
The subscripts $f$ and $s$ indicate the
fastest and slowest particles in the laboratory respectively
and $X^0$ represents the central
system that is presumed to be produced by double exchange processes.
The experiment
has been performed using the CERN Omega Spectrometer,
the layout of which is
described in ref.~\cite{WADPT}.
\par
Analyses of the \kkpi and \pipipipi
channels
have been published
previously~\cite{kkpi,4pi}.
This paper presents a study of the
$f_1(1285)$ and $f_1(1420)$
through an analysis of the
\etapipi, \rhog and \phig channels.
\par
The reaction
\begin{equation}
pp \rightarrow p_{f} (\eta\pi^+\pi^-) p_{s}
\label{eq:c}
\end{equation}
where the $\eta$ has been observed decaying to $\gamma \gamma$ and
$\pi^+\pi^-\pi^0$,
has been isolated
from the sample of events having four and six
outgoing
charged tracks respectively
plus two $\gamma$s each with energy greater than 0.5 GeV
reconstructed in the electromagnetic
calorimeter\footnote{The showers associated with the impact of
the charged tracks on the calorimeter
have been removed from the event before the
requirement of only two $\gamma$s was made.},
by first imposing the following cuts on the components of the
missing momentum:
$|$missing~$P_{x}| <  17.0$ GeV/c,
$|$missing~$P_{y}| <  0.16$ GeV/c and
$|$missing~$P_{z}| <  0.12$ GeV/c,
where the x axis is along the beam
direction.
A correlation between
pulse-height and momentum
obtained from a system of
scintillation counters was used to ensure that the slow
particle was a proton.
\par
For the case $\eta \rightarrow \gamma \gamma$
the effective mass of the two $\gamma$s
is shown in fig.~\ref{fi:1}a) for the events contributing to the
$f_1(1285)$ mass region of the
\etapipi spectrum.
The $\gamma \gamma$ mass spectrum
shows
a clear $\eta$ signal ($\sigma$~=~31~MeV) with little background,
which was selected by requiring
0.45~$<$~m($\gamma \gamma$)~$<$~0.65~GeV.
For the case $\eta \rightarrow \pi^+\pi^-\pi^0$
the effective mass of the \pipipi
is shown in fig.~\ref{fi:1}b) for the events contributing to the
$f_1(1285)$ mass region of the
\etapipi spectrum.
The \pipipi mass spectrum
shows
a clear $\eta$ signal ($\sigma$~=~11~MeV) with effectively no background,
which was selected by requiring
0.5~$<$~m($\pi^+ \pi^- \pi^0$)~$<$~0.6~GeV.
The quantity $\Delta$, defined as
$ \Delta = MM^{2}(p_{f}p_{s}) - M^{2}(\eta \pi^{+}\pi^{-})$,
where $MM^{2}(p_{f}p_{s})$ is the missing mass squared of the two
outgoing protons,
was then calculated for each event and
a cut of $|\Delta|$ $\leq$ 3.0 GeV$^{2}$ was used to select the
$\eta\pi^{+}\pi^{-}$
channel. Events containing a fast $\Delta^{++}(1232) $
were removed if $M(p_{f} \pi^{+}) < 1.3 $ GeV, which left
90~393 centrally produced \etapipi events for $\eta \rightarrow \gamma \gamma$
and 24~585 for $\eta \rightarrow \pi^+\pi^-\pi^0$.
After the selection of the \etapipi channel a kinematical fit was performed in
order to apply overall energy and momentum balance.
\par
Fig.~\ref{fi:1}c) and d) shows the
$\eta \pi^{+}\pi^{-}$
effective mass spectrum for $\eta \rightarrow \gamma \gamma$ and
$\eta \rightarrow \pi^+\pi^-\pi^0$ respectively.
Clear peaks of the $\eta^\prime$ and $f_1(1285)$ can be seen
together with a shoulder in the 1.4~GeV mass region.
A fit to these spectra has been attempted
using a Gaussian to describe the $\eta^\prime$, a
Breit-Wigner convoluted with a Gaussian, to
account for the experimental resolution
($\sigma$~=~21~MeV for $\eta \rightarrow \gamma \gamma$ and
$\sigma$~=~17~MeV for $\eta \rightarrow \pi^+ \pi^- \pi^0$),
to describe the $f_1(1285)$
and a background
of the form $a(m-m_{th})^b exp(-cm-dm^2)$, where
$m$ is the \etapipi mass,
$m_{th}$ is the threshold mass and
$a$, $b$, $c$ ,$d$ are fit parameters. The fit (not shown) fails
to describe the 1.4~GeV region.
Fig.~\ref{fi:2}a) shows the result of the fit
(for the case $\eta \rightarrow \gamma \gamma$)
including a Breit-Wigner to describe the 1.4 GeV region
which yields a mass of 1370~$\pm$~5~MeV and a width of 109~$\pm$~18~MeV.
We did not observe anything in this mass region in the analysis of the
\kkpi channel~\cite{kkpi} and these
parameters are lower in mass and broader in width than states
observed in the \etapipi channel by other experiments~\cite{PDG98}.
This may suggest that the shoulder is due to an interference effect and
this possibility will be discussed below.
\par
A spin-parity analysis of the
\etapipi
channel
has been performed in the mass interval 1.2 to 1.5 GeV
using an isobar model~\cite{re:wa914pi}.
Assuming that
only angular momenta up to 2 contribute,
the intermediate
states considered are $a_0(980) \pi$ and $\sigma \eta$,
where $\sigma$ stands for the low mass $\pi\pi$ S-wave amplitude
squared~\cite{re:zbugg}.
The amplitudes have been calculated in the spin-orbit (LS)
scheme using spherical
harmonics.
In order to perform a spin parity analysis the
log likelihood function, ${\cal L}_{j}=\sum_{i}\log P_{j}(i)$,
is defined by combining the probabilities of all events in 20 MeV
\etapipi mass bins from 1.2 to 1.5 GeV.
The incoherent sum of various
event fractions $a_{j}$ is calculated
so as to include more than one wave in the fit,
\begin{equation}
{\cal L}=\sum_{i}\log \left(\sum_{j}a_{j}P_{j}(i) +
(1-\sum_{j}a_{j})\right)
\end{equation}
where the term
$(1-\sum_{j}a_{j})$ represents the phase space background.
The negative log likelihood function ($-{\cal L} $) is then minimised using
MINUIT~\cite{re:MINUIT}.
Different combinations of waves and isobars have been tried and
insignificant contributions have been removed from the final fit.
\par
The only wave required in the fit is the $J^{PC}$~=~$1^{++}$ $a_0(980)\pi$
wave with spin projection $|J_z|$~=~1.
No $J^{PC}$~=~$0^{-+}$ $a_0(980)\pi$ or any $\sigma \eta$ waves
are required in the fit.
Fig.~\ref{fi:2}c) shows the
$J^{PC}$~=~$1^{++}$ $a_0(980)\pi$ wave
where the $f_1(1285)$ and the shoulder
at 1.4~GeV can be seen.
Even though the addition of the
$J^{PC}$~=~$0^{-+}$ $a_0(980)\pi$ wave makes an insignificant change to the
likelihood we have included it and show the result in
fig.~\ref{fi:2}d) where a flat distribution can be observed.
\par
The fact that the shoulder at 1.4~GeV appears in the
$J^{PC}$~=~$1^{++}$ $a_0(980)\pi$ wave suggests
that it could be due to the $f_1(1420)$.
However, the mass obtained using
two incoherent Breit-Wigners, as described above, did not
give mass and width parameters compatible with the $f_1(1420)$.
If the $f_1(1420)$ did have an $a_0(980)\pi$ decay mode then
a possible solution is that,
since both the $f_1(1285)$ and $f_1(1420)$
have the same quantum numbers and decay mode, they could interfere.
\par
In order to test the interference hypothesis
a fit has been performed to the \etapipi mass spectrum
(shown in fig.~\ref{fi:2}b))
and to the
$J^{PC}$~=~$1^{++}$ $a_0(980)\pi$ wave
in fig.~\ref{fi:2}c) using a K matrix formalism~\cite{KMATRIX}
including poles to describe the interference between the
$f_1(1285)$ and the $f_1(1420)$. The parameters of the poles have been
fixed to the values obtained for the
$f_1(1285)$ and the $f_1(1420)$ from a fit to the \kkpi mass
spectrum~\cite{kkpi}
and the resulting function smeared to take into account the experimental
resolution.
The only free parameters are the relative phase between the resonances
and their relative production strengths.
As can be seen from
fig.~\ref{fi:2}b) and c)
the parameterisation describes well both the total mass spectrum and the
$J^{PC}$~=~$1^{++}$ $a_0(980)\pi$ wave. The relative phase is found to be
154~$\pm$~9~degrees.
Therefore, the shoulder at 1.4 GeV can be interpreted as an $a_0(980)\pi$
decay mode of the $f_1(1420)$.
\par
In order to determine the branching ratio of the $f_1(1285)$ and the
$f_1(1420)$ a number of checks have been employed.
Firstly, we have measured the acceptance corrected number of
$\eta^\prime$ and $f_1(1285)$ observed in the \etapipi mass spectrum
for the cases where the $\eta$ decays to $\gamma \gamma$ and
\pipipi and hence have determined the $\eta$ branching ratio to be
\begin{equation}
\frac{\eta \rightarrow \gamma \gamma}{\eta \rightarrow \pi^+\pi^-\pi^0} = 1.70
\pm 0.09 \pm 0.02
\end{equation}
which is in good agreement with the PDG value of 1.70~$\pm$~0.04~\cite{PDG98}.
\par
We have previously measured the $f_1(1285)$ branching ratio in the
\pipipipi and \kkpi channels. Here we compare the $\eta\pi\pi$ to the
$K \overline K \pi$ for which we determine
\begin{equation}
\frac{f_1(1285) \rightarrow K \overline K \pi}{f_1(1285) \rightarrow \eta \pi
\pi} = 0.166 \pm 0.01 \pm 0.008
\end{equation}
which is in agreement with the PDG value of 0.19~$\pm$~0.07~\cite{PDG98}
and represents an appreciable improvement in precision.
\par
The $f_1(1420)$ has also been observed in the $K^*(892) \overline K$
decay mode by this experiment~\cite{kkpi} and hence its
branching ratio to $a_0(980)\pi$ and
$K^*(892) \overline K$
has been calculated taking into account the unseen decay modes and geometrical
acceptance effects
and gives
\begin{equation}
\frac{f_1(1420) \rightarrow a_0(980) \pi}{f_1(1420) \rightarrow K^*(892)
\overline K} = 0.04 \pm 0.01 \pm 0.01.
\end{equation}
Hence a small ($\approx 4$ \%) contribution for the $a_0(980)\pi$ decay mode
should have been found in the analysis of the \kkpi channel. However,
a contribution at this level would have been at the limit of
sensitivity.
\par
There is considerable uncertainty in the branching ratio of the
$a_0(980)$~\cite{PDG98}.
Since the $f_1(1285)$ is observed
in both the \kkpi and \etapipi channels
decaying effectively 100~\% to
$a_0(980)\pi$
we can use this information to determine the branching ratio of the $a_0(980)$
to be
\begin{equation}
\frac{a_0(980) \rightarrow K \overline K}{a_0(980) \rightarrow \eta \pi} =
0.166 \pm 0.01 \pm 0.02
\end{equation}
where the systematic error has been increased to take into account any
possible non-$a_0(980)\pi$ decay of the $f_1(1285)$.
\par
We have previously published a paper on the \rhog final state~\cite{oldrhog}
in which we showed that there was only evidence for the $\eta^\prime$ and
$f_1(1285)$. Using the increase of a factor of ten in statistics
in this current paper we want to improve on the measurement of the $f_1(1285)$
branching ratio and the upper limit for $f_1(1420)$~$\rightarrow$~\rhog.
In addition, we want to search for possible \phig decay modes of both
resonances.
The reaction
\begin{equation}
pp \rightarrow p_{f} (h^+h^-\gamma) p_{s}
\label{eq:d}
\end{equation}
has been isolated
from the sample of events having four
outgoing
charged tracks plus one $\gamma$ with energy greater than 2.0 GeV
reconstructed in the electromagnetic
calorimeter
by first imposing the following cuts on the components of the
missing momentum:
$|$missing~$P_{x}| <  12.0$ GeV/c,
$|$missing~$P_{y}| <  0.10$ GeV/c and
$|$missing~$P_{z}| <  0.06$ GeV/c.
For the
\kkg channel one of the central
particles
is required to be
identified as being a $K$ by the {\v C}erenkov system and the other particle
is required to be compatible with being a $K$.
Energy balance is then used to select out the \pipig and \kkg channels.
The major backgrounds to these channels come from the more
copiously produced \pipipi and
$K^+K^-\pi^0$ channels where one of the $\gamma$s from the decay of the
$\pi^0$ is not detected and the resulting event passes
through the momentum balance cuts.
The contribution of this background to the selected channels has
been simulated taking real \pipipi and $K^+K^-\pi^0$ events
and distributing them according to a Gaussian in Feynman x ($x_F$)
centered at $x_F$~=~0
in which the $\pi^0$ has been allowed to decay isotropically.
\par
Fig.~\ref{fi:3}a) shows the selected \pipig mass spectrum together
with the expected background from the \pipipi channel.
Fig.~\ref{fi:3}b) shows the resulting background subtracted \pipig mass
spectrum where evidence can be seen for the $\eta$, $\eta^\prime$ and
$f_1(1285)$.
Fig.~\ref{fi:3}c) shows the \pipi mass spectrum from the selected \pipig
events with the expected background from the \pipipi channel and
fig.~\ref{fi:3}d) shows the resulting background subtracted mass
spectrum which shows a threshold enhancement due to the
$\eta$~$\rightarrow$~\pipig and a signal which has been identified as
the $\rho^0(770)$.
The \rhog channel has been selected by requiring
0.70~$<$~m($\pi^+\pi^-$)~$<$~0.84~GeV.
Fig.~\ref{fi:3}e) shows the \rhog mass spectrum from the selected \pipig
events with the expected background from the \pipipi channel and
fig.~\ref{fi:3}f) shows the resulting background subtracted mass
spectrum which shows clear $\eta^\prime$ and $f_1(1285)$ signals. There
is no evidence for any signal in the 1.4~GeV mass region.
\par
After acceptance correcting the data and taking into account the
unseen decay modes we have determined the branching ratios of the
$\eta^\prime$ and $f_1(1285)$ to be
\begin{equation}
\frac{\eta^\prime \rightarrow \rho^0 \gamma}{\eta^\prime \rightarrow \eta
\pi\pi} = 0.43 \pm 0.02 \pm 0.02
\end{equation}
and
\begin{equation}
\frac{f_1(1285) \rightarrow \rho^0 \gamma}{f_1(1285) \rightarrow \eta \pi\pi} =
0.10 \pm 0.01 \pm 0.02
\end{equation}
both of which are in good agreement with the PDG values of 0.459~$\pm$~0.025
and 0.108~$\pm$~0.046 respectively~\cite{PDG98}.
\par
For the $f_1(1420)$ an upper limit has been calculated to be
\begin{equation}
\frac{f_1(1420) \rightarrow \rho^0 \gamma}{f_1(1420) \rightarrow K \overline K
\pi} < 0.02\thinspace\thinspace ( 95 \% \thinspace \thinspace c.l.).
\end{equation}
This is an improvement on our previous upper limit of $<$~0.08~\cite{oldrhog}.
\par
For the \phig channel there is an additional problem in that the signals
that we wish to study, namely the $f_1(1285)$ and the $f_1(1420)$, are also
present in the background $K^+K^-\pi^0$ channel.
However, since the method of background subtraction has been successful
for the \pipig channel the same method will be employed here.
Fig.~\ref{fi:4}a) shows the selected \kkg mass spectrum together
with the expected background from the $K^+K^-\pi^0$ channel.
Fig.~\ref{fi:4}b) shows the resulting background subtracted \kkg mass
spectrum.
Fig.~\ref{fi:4}c) shows the $K^+K^-$ mass spectrum from the selected \kkg
events with the expected background from the $K^+K^-\pi^0$ channel and
fig.~\ref{fi:4}d) shows the resulting background subtracted mass
spectrum which shows evidence for a $\phi$ signal.
The \phig channel has been selected by requiring
1.01~$<$~m($K^+K^-$)~$<$~1.03~GeV.
Fig.~\ref{fi:4}e) shows the \phig mass spectrum from the selected \kkg
events with the expected background from the $K^+K^-\pi^0$ channel and
fig.~\ref{fi:4}f) shows the resulting background subtracted mass
spectrum which shows a possible excess of events
in the $f_1(1420)$ region.
\par
If the events in the 1.42~GeV mass region are interpreted as being due
to the $f_1(1420)$ then
after acceptance correcting the data and taking into account the
unseen decay modes we have determined the branching ratio of the
$f_1(1420)$ to be
\begin{equation}
\frac{f_1(1420) \rightarrow \phi \gamma}{f_1(1420) \rightarrow K \overline K
\pi} = 0.003 \pm 0.001 \pm 0.001.
\end{equation}
For the $f_1(1285)$ we can only calculate an upper limit of
\begin{equation}
\frac{f_1(1285) \rightarrow \phi \gamma}{f_1(1285) \rightarrow K \overline K
\pi} < 0.005 \thinspace\thinspace  (95 \% \thinspace \thinspace c.l.).
\end{equation}
The number obtained for the $f_1(1285)$ is compatible with the
PDG value of 0.0082~$\pm$~0.0033~$\pm$~0.0020. A value for the $f_1(1420)$
has not previously been measured.
\par
Assuming that the only decay modes of the $f_1(1285)$ are $\pi\pi\pi\pi$,
$K \overline K \pi$, $\eta \pi\pi$, \rhog and \phig we have determined the
branching fractions of the $f_1(1285)$ as shown in table~\ref{ta:a}.
These values are compatible with those found in the PDG~\cite{PDG98}
but represent an increase in precision.
Assuming that the only decay modes of the $f_1(1420)$ are
$K^*(892) \overline K$,
$a_0(980) \pi$ and \phig we have determined the
branching fractions of the $f_1(1420)$ as shown in table~\ref{ta:b}.
\par
We have previously published the $dP_T$ dependence,
where $dP_T$ is the difference
in the transverse momentum vectors of the two exchange
particles~\cite{WADPT,closeak},
of the $f_1(1285)$~\cite{kkpi,4pi} and $f_1(1420)$~\cite{kkpi}.
However,
due to an improved understanding and simulation of the experimental trigger,
these values have now changed from the previously published
values~\cite{kkpi,4pi}.
The fraction of $f_1(1285)$ and $f_1(1420)$ has
been calculated for
$dP_T$$\leq$0.2 GeV, 0.2$\leq$$dP_T$$\leq$0.5 GeV and $dP_T$$\geq$0.5 GeV
and gives
0.03~$\pm$~0.01, 0.35~$\pm$~0.02 and 0.61~$\pm$~0.04 for the $f_1(1285)$ and
0.02~$\pm$~0.02, 0.38~$\pm$~0.02 and 0.60~$\pm$~0.04 for the $f_1(1420)$.
This results in a ratio of production at small $dP_T$ to large $dP_T$ of
0.05~$\pm$~0.016 for the $f_1(1285)$ and
0.03~$\pm$~0.03 for the $f_1(1420)$.
These ratios are similar to those found for other $q \overline q$
mesons~\cite{memoriam}.
\par
In order to determine the
four momentum transfer dependence ($t$) of the $f_1(1285)$ and $f_1(1420)$
the \etapipi and \kkpi mass spectrum have been fitted in 0.1 GeV$^2$ bins
of $t$.
Fig.~\ref{fi:5}a) and b) show the four momentum transfer from
one of the proton vertices
for the $f_1(1285)$ and $f_1(1420)$ respectively.
The distributions have been fitted with a single exponential
of the form $exp(-b |t|)$ and yields $b$~=~6.3~$\pm$~0.3~GeV$^{-2}$ for the
$f_1(1285)$ and
5.6~$\pm$~0.5~GeV$^{-2}$ for the
$f_1(1420)$.
These values
are consistent
with what is expected from Double Pomeron Exchange (DPE)~\cite{dpet}.
\par
The azimuthal angle ($\phi$) is defined as the angle between the $p_T$
vectors of the two protons. The $\phi$ dependence of the
$f_1(1285)$ and $f_1(1420)$ has been determined by fitting
the \etapipi and \kkpi mass spectrum in 20 degree bins in $\phi$.
The resulting distributions are shown in
fig.~\ref{fi:5}c) and d)
for the $f_1(1285)$ and $f_1(1420)$ respectively.
These distributions are compatible with each other but
differ significantly from those observed
for the $\pi^0$, $\eta$, $\eta^\prime$~\cite{0mpap},
$\omega$~\cite{3pipap} and the $\phi \phi$~\cite{phiphi} and
\kstkst~\cite{kstkst} final states.
\par
After correcting for
geometrical acceptances, detector efficiencies,
losses due to cuts,
and unseen decay modes,
the cross-section for
$f_1(1285)$ and $f_1(1420)$ production at $\sqrt s$~=~29.1~GeV in the
$x_F$ interval
$|x_F| \leq 0.2$ is $\sigma(f_1(1285))$~=~6919~$\pm$~886~nb
and $\sigma(f_1(1420))$~=~1584~$\pm$~145~nb.
This can be compared with the cross-sections found in the same interval
at $\sqrt s$~=~12.7~\cite{cenkkpi} which we have recalculated to be
$\sigma(f_1(1285))$~=~6857~$\pm$~1306~nb
and $\sigma(f_1(1420))$~=~1080~$\pm$~385~nb.
The cross section as a function of energy for
both the $f_1(1285)$ and $f_1(1420)$ is found to be consistent with
being flat or rising slightly with centre of mass energy which
is consistent with them being produced via DPE~\cite{dpet}.
In a previous publication it had been observed~\cite{cenkkpi} that
the cross-section  for the $f_1(1285)$ decreased from
$\sqrt s$~=~12.7 to 23.8~GeV. However, this was due to the fact that
the experiment at $\sqrt s$~=~23.8~GeV was only sensitive to $\phi$ angles
less than 90 degrees and
the acceptance program that had been used assumed a flat $\phi$ distribution.
Hence it underestimated the cross section.
\par
In conclusion,
central production is a clean process in which to study the properties
of the $f_1(1285)$ and $f_1(1420)$ due to their strong production
relative to $J^{PC}$~=~$0^{-+}$ states in the same mass region.
The branching fractions of the $f_1(1285)$ and $f_1(1420)$
in all major decay modes
have been determined with improved accuracy.
Both the $f_1(1285)$ and $f_1(1420)$ are consistent with being produced
by DPE.
\begin{center}
{\bf Acknowledgements}
\end{center}
\par
This work is supported, in part, by grants from
the British Particle Physics and Astronomy Research Council,
the British Royal Society,
the Ministry of Education, Science, Sports and Culture of Japan
(grants no. 04044159 and 07044098), the Programme International
de Cooperation Scientifique (grant no. 576)
and
the Russian Foundation for Basic Research
(grants 96-15-96633 and 98-02-22032).

\newpage
\newpage
\begin{table}[h]
\caption{The branching fractions of the $f_1(1285)$.}
\label{ta:a}
\vspace{0.5in}
\begin{center}
\begin{tabular}{|c|c|} \hline
  & \\
Decay Mode & Fraction (\%) \\
 &\\ \hline
 & \\
$\pi\pi\pi\pi$  & $33.0 \pm 1.5 \pm 1.5$  \\
 & \\
$K \overline K \pi$  & $8.7 \pm 0.4 \pm 0.4$  \\
 & \\
$\eta\pi\pi$  & $52.8 \pm 3.8 \pm 2.5$  \\
 & \\
$\rho^0 \gamma$  & $5.4 \pm 0.4 \pm 0.3$  \\
 & \\
$\phi \gamma$  & $< 0.043 \thinspace \thinspace (95 \% \thinspace \thinspace
c.l.)$  \\
  & \\ \hline
\end{tabular}
\end{center}
\end{table}
\begin{table}[h]
\caption{The branching fractions of the $f_1(1420)$.}
\label{ta:b}
\vspace{0.5in}
\begin{center}
\begin{tabular}{|c|c|} \hline
  & \\
Decay Mode & Fraction (\%) \\
 &\\ \hline
 & \\
$K^*(892) \overline K$  & $96.0 \pm 1.0 \pm 1.0$  \\
 & \\
$a_0(980) \pi$  & $4.0 \pm 1.0 \pm 1.0$  \\
 & \\
$\phi \gamma$  & $0.3 \pm 0.12 \pm 0.2$  \\
  & \\ \hline
\end{tabular}
\end{center}
\end{table}
\newpage
{ \large \bf Figures \rm}
\begin{figure}[h]
\caption{
a) The $\gamma \gamma$ mass spectrum and b) the \pipipi mass spectrum
contributing to the
$f_1(1285)$ mass region of the \etapipi spectrum.
The \etapipi mass spectrum for c) $\eta \rightarrow \gamma \gamma$ and
d) $\eta \rightarrow \pi^+\pi^-\pi^0$.
}
\label{fi:1}
\end{figure}
\begin{figure}[h]
\caption{
a) and b) the \etapipi mass spectrum with fits described in the text.
c) The $J^{PC}~=~1^{++}$~$a_0(980)\pi$ wave and
d) the $J^{PC}~=~0^{-+}$~$a_0(980)\pi$ wave.
}
\label{fi:2}
\end{figure}
\begin{figure}[h]
\caption{a) The \pipig mass spectrum with shaded the background from the
\pipipi channel b) the background subtracted \pipig mass spectrum.
c) The \pipi mass spectrum with shaded the background from the
\pipipi channel d) the background subtracted \pipi mass spectrum.
e) The \rhog mass spectrum with shaded the background from the
\pipipi channel f) the background subtracted \rhog mass spectrum.
}
\label{fi:3}
\end{figure}
\begin{figure}[h]
\caption{a) The \kkg mass spectrum with shaded the background from the
$K^+K^-\pi^0$ channel b) the background subtracted \kkg mass spectrum.
c) The $K^+K^-$ mass spectrum with shaded the background from the
$K^+K^-\pi^0$ channel d) the background subtracted $K^+K^-$ mass spectrum.
e) The \phig mass spectrum with shaded the background from the
$K^+K^-\pi^0$ channel f) the background subtracted \phig mass spectrum.
}
\label{fi:4}
\end{figure}
\begin{figure}[h]
\caption{The four momentum transfer squared ($|t|$) from one of the proton
vertices for a) the $f_1(1285)$ and b) the $f_1(1420)$.
The azimuthal angle ($\phi$) between the two outgoing protons
for c) the $f_1(1285)$ and d) the $f_1(1420)$.
}
\label{fi:5}
\end{figure}
\newpage
\begin{center}
\epsfig{figure=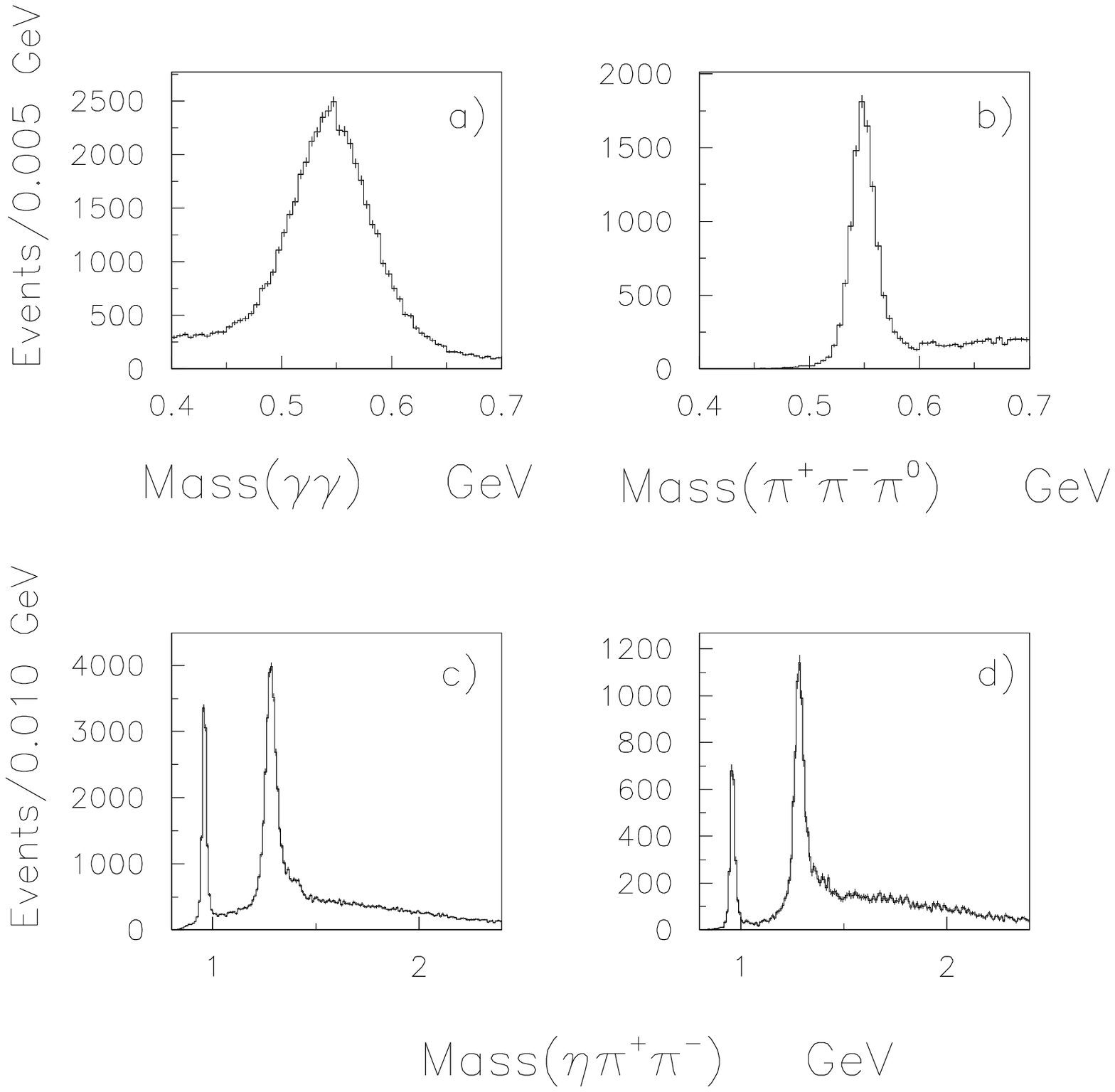,height=22cm,width=17cm}
\end{center}
\begin{center} {Figure 1} \end{center}
\newpage
\begin{center}
\epsfig{figure=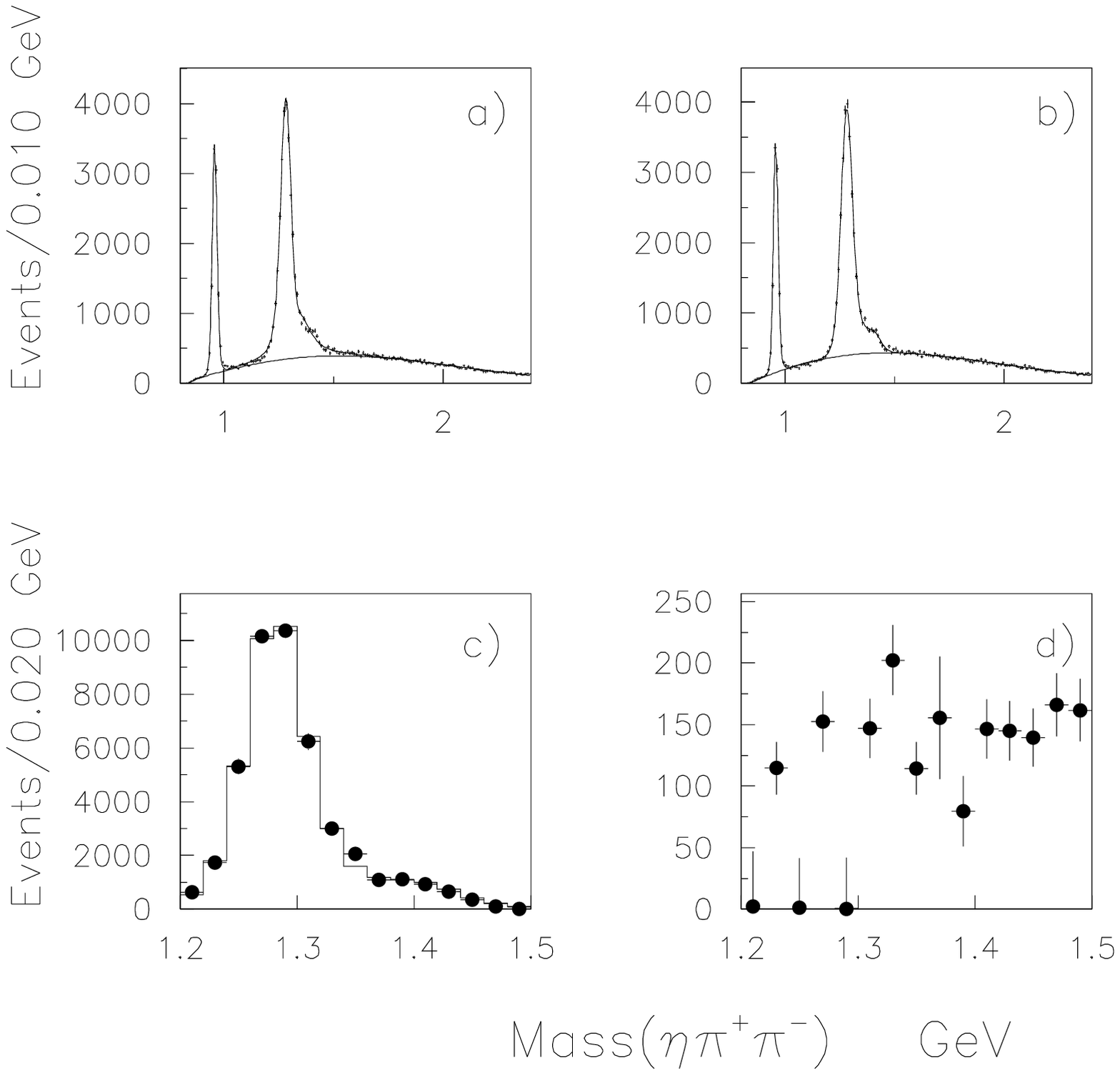,height=22cm,width=17cm}
\end{center}
\begin{center} {Figure 2} \end{center}
\newpage
\begin{center}
\epsfig{figure=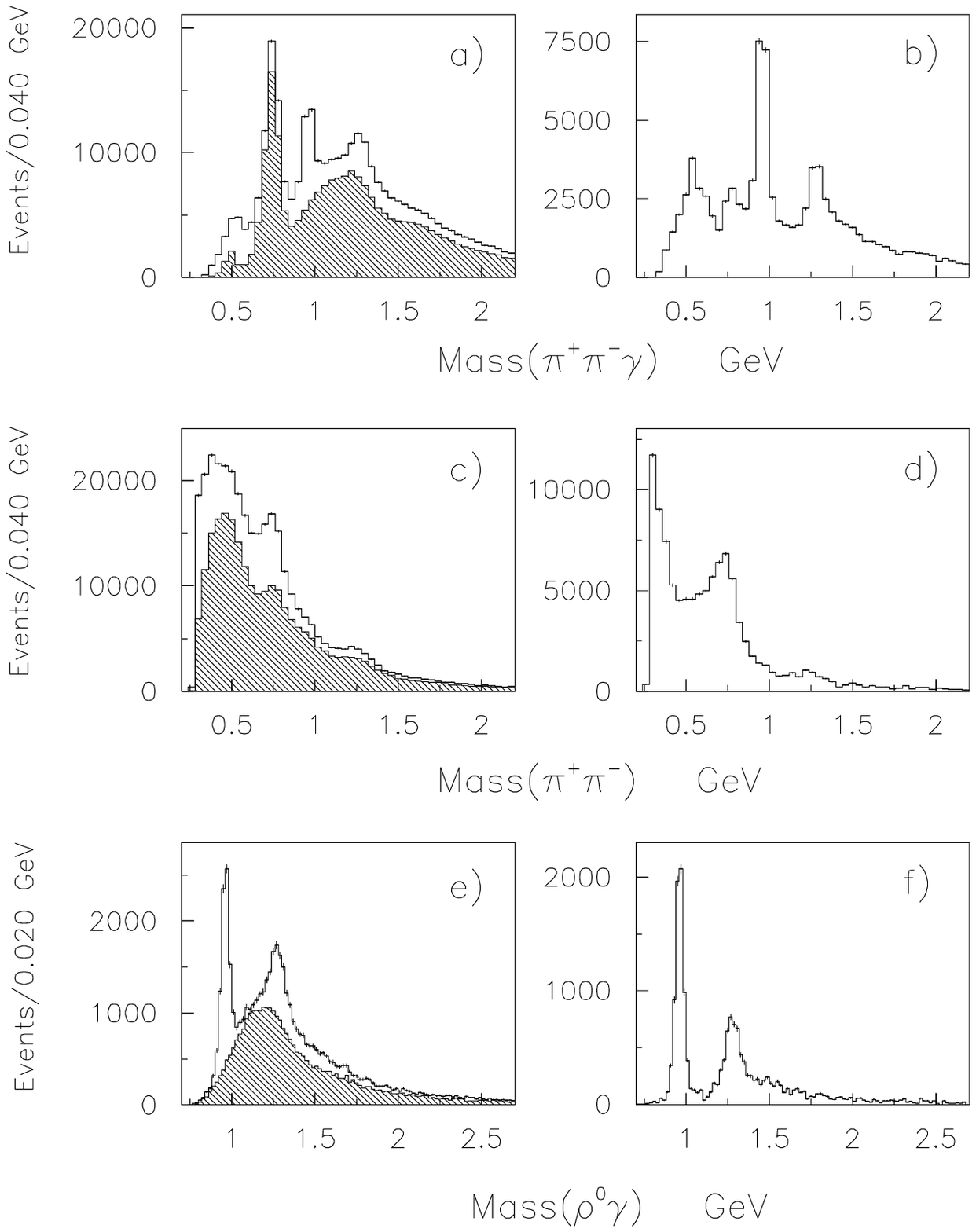,height=22cm,width=17cm}
\end{center}
\begin{center} {Figure 3} \end{center}
\newpage
\begin{center}
\epsfig{figure=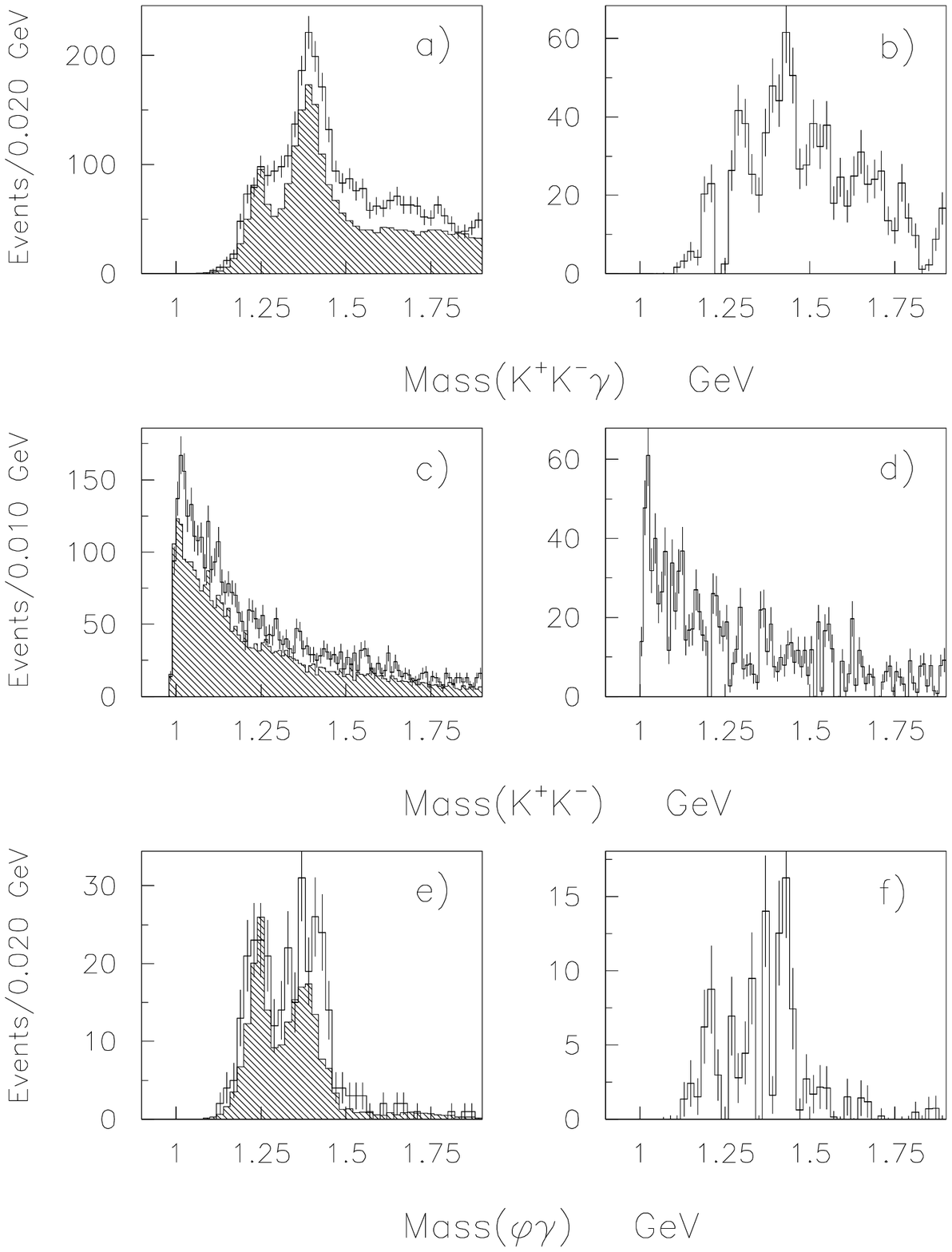,height=22cm,width=17cm}
\end{center}
\begin{center} {Figure 4} \end{center}
\newpage
\begin{center}
\epsfig{figure=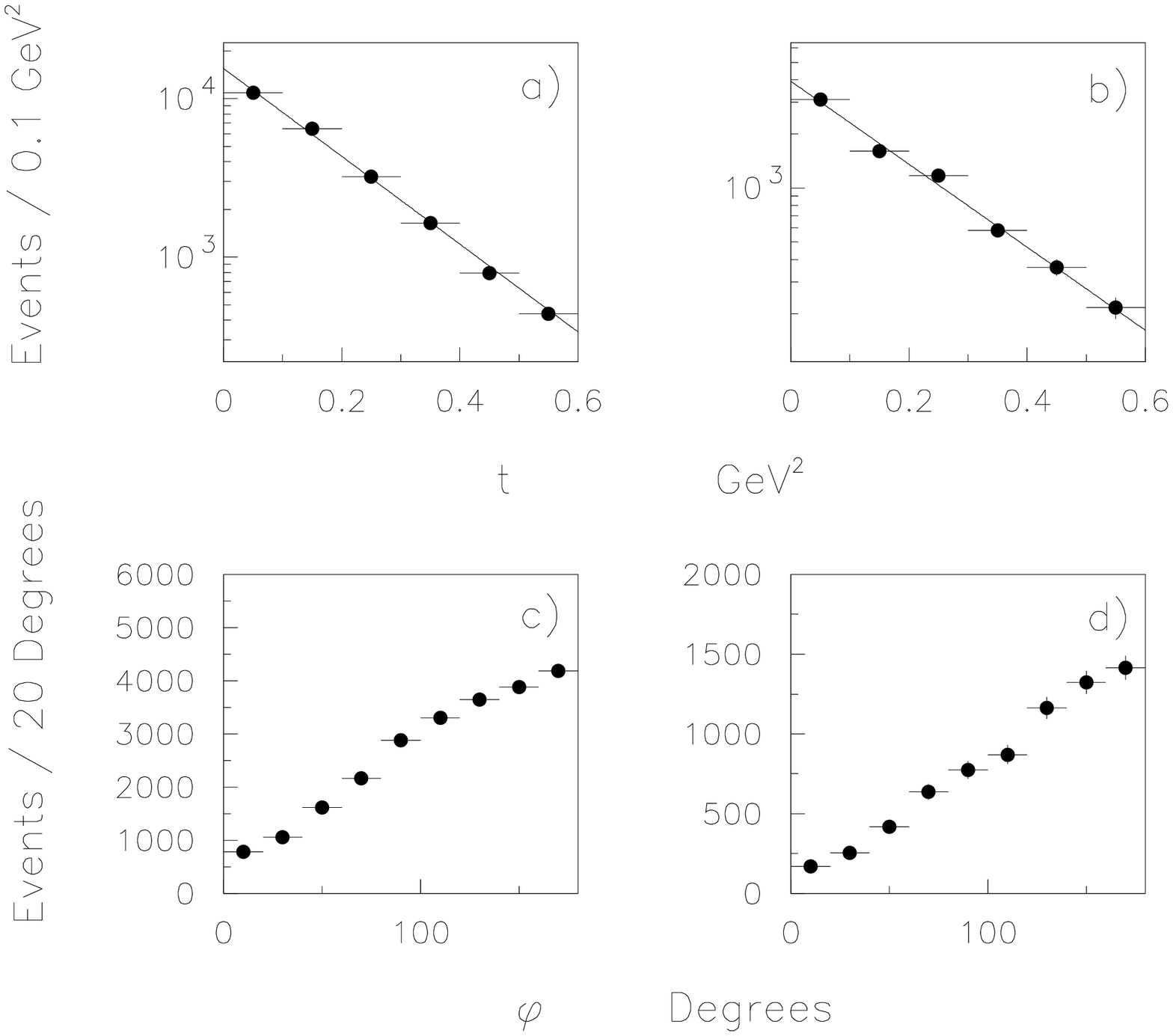,height=22cm,width=17cm}
\end{center}
\begin{center} {Figure 5} \end{center}

\bigskip
\newpage
\begin{thebibliography}{99}
\bibitem{wa76kkpi}
T. Armstrong {\em et al.,} Phys. Lett. {\bf B146} (1984) 273; \\
\bibitem{cenkkpi}
T. Armstrong {\em et al.,} Zeit. Phys. {\bf C34} (1987) 23; \\
T. Armstrong {\em et al.,} Phys. Lett. {\bf B221} (1989) 216; \\
T. Armstrong {\em et al.,} Zeit. Phys. {\bf C56} (1992) 29.
\bibitem{cenetapipi}
T. Armstrong {\em et al.,} Zeit. Phys. {\bf C52} (1991) 389.
\bibitem{E690}
M.C. Berisso {\em et al.,} A.I.P. conf. proc. {\bf 432} (1997) 36.
\bibitem{WADPT}
D. Barberis {\em et al.,} Phys. Lett. {\bf B397 } \rm (1997) 339.
\bibitem{kkpi}
D. Barberis {\em et al.,} Phys. Lett. {\bf B413} \rm (1997) 225.
\bibitem{4pi}
D. Barberis {\em et al.,} Phys. Lett. {\bf B413} \rm (1997) 217.
\bibitem{PDG98}
Particle Data Group, European Physical Journal {\bf C3} (1998) 1.
\bibitem{re:wa914pi}
S. Abatzis {\em et al.,} Phys. Lett. {\bf B324 } \rm (1994) 509.
\bibitem{re:zbugg}
B. S. Zou and D. V. Bugg, Phys. Rev. {\bf D48} (1993) R3948.
\bibitem{re:MINUIT}
F. James and M. Roos, MINUIT Computer Physics Communications
{\bf 10 } \rm (1975) 343; CERN-D506 (1989).
\bibitem{KMATRIX}
S.U. Chung {\em et al.,} Ann. d. Physik. {\bf 4} (1995) 404.
\bibitem{oldrhog}
T. Armstrong {\em et al.,} Zeit. Phys. {\bf C54} (1992) 371.
\bibitem{closeak}
F.E. Close and A. Kirk, Phys. Lett. {\bf B397 } \rm (1997) 333.
\bibitem{memoriam}
A. Kirk, hep-ex/9803024.
\bibitem{dpet}
S.N. Ganguli and D.P. Roy, Phys. Rep. {\bf 67} (1980) 203.
\bibitem{0mpap}
D. Barberis {\em et al.,} Phys. Lett. {\bf B427} (1998) 398.
\bibitem{3pipap}
D. Barberis {\em et al.,} Phys. Lett. {\bf B422} (1998) 399.
\bibitem{phiphi}
D. Barberis {\em et al.,} Phys. Lett. {\bf B432} (1998) 436.
\bibitem{kstkst}
D. Barberis {\em et al.,} hep-ex/9807021.
\end{thebibliography}
\end{document}